\documentclass[preprint,12pt]{elsarticle}

\usepackage{amsmath,amssymb,amsfonts,amsthm}
\usepackage{bm}
\usepackage{mathtools}
\usepackage{graphicx}
\usepackage{epsfig,epstopdf}
\usepackage{subfig}
\usepackage{nicefrac}
\usepackage{natbib}
\usepackage{tabularx}
\usepackage{multirow}
\usepackage{xcolor}
\usepackage{hyperref}
\usepackage{makecell}

\hypersetup{
    colorlinks=true,
    linkcolor=blue,
    citecolor=blue,
    urlcolor=blue
}

\usepackage{algorithm}
\usepackage{algpseudocode}

\usepackage{tikz}
\usetikzlibrary{calc}

\theoremstyle{definition}

\newcommand{\wb}{{\textbf{w}}}
\newcommand{\xb}{{\textbf{x}}}
\newcommand{\yb}{{\textbf{y}}}

\newcommand{\Sbb}{{\textbf{S}}}

\newcommand{\nb}{{\textbf{n}}}
\newcommand{\Nb}{{\textbf{N}}}
\newcommand{\Ib}{{\textbf{I}}}

\newcommand{\vb}{{\textbf{v}}}

\newcommand{\Hb}{\mathbf{H}}

\newcommand{\Rb}{\mathbf{R}}
\newcommand{\Wb}{\mathbf{W}}

\newcommand{\Yb}{\mathbf{Y}}
\newcommand{\Zb}{\mathbf{Z}}

\begin{document}

\begin{frontmatter}

\title{Blind Channel and Noise-Variance Estimation for MIMO Thermal Noise Modulation Systems via Generalized Expectation-Maximization}

\author[basu]{Razieh Torkamani\corref{cor1}}
\ead{r.torkamani@basu.ac.ir}

\author[inatel]{Hadi Zayyani}
\ead{hadi.zayyan@posdoc.inatel.br}

\author[aum]{Mohammad Salman}
\ead{mohammad.salman@aum.edu.kw}

\author[inatel]{Felipe A. P. de Figueiredo}
\ead{felipe.figueiredo@inatel.br}

\cortext[cor1]{Corresponding author}

\address[basu]{Department of Electrical Engineering, Faculty of Engineering,
Bu-Ali Sina University, Hamedan, Iran}

\address[inatel]{National Institute of Telecommunications (Inatel),
Santa Rita do Sapucaí, Brazil}

\address[aum]{College of Engineering and Technology,
American University of the Middle East, Egaila, 54200, Kuwait}

\begin{abstract}
Thermal Noise Modulation (TNM) has recently emerged as a communication paradigm in which information is conveyed through different thermal-noise variance levels rather than deterministic signal waveforms. The stochastic nature of TNM signals poses fundamental challenges for conventional pilot-based channel estimation, while pilot transmission introduces additional overhead and reduces spectral efficiency. This paper proposes a pilot-free blind receiver for M-ary TNM systems based on the Generalized Expectation-Maximization (GEM) framework. The receiver jointly estimates the channel Gram matrix, transmitted variance states, mixture probabilities, and unknown transmit variance levels directly from the received observations without requiring pilot symbols or prior knowledge of the transmit variance constellation. The modulation order $M$ and the receiver-noise variance $\sigma_w^2$ are assumed to be known at the receiver. The proposed receiver jointly estimates the channel matrix, latent transmitted symbols, mixture probabilities, and unknown transmit-noise variance levels directly from the received observations, without requiring pilot symbols or prior knowledge of the variance constellation. To address the absence of a closed-form solution for the maximization step, a gradient-ascent procedure is incorporated into the GEM framework to iteratively update the unknown parameters. The proposed approach is applicable to general $M$-ary TNM signaling and enables simultaneous channel estimation, symbol inference, and variance learning within a unified blind-receiver framework. Simulation results demonstrate accurate estimation of the channel Gram matrix and effective symbol detection while completely eliminating pilot overhead. Furthermore, the proposed variance-learning scheme achieves performance close to that obtained when the transmit variance levels are perfectly known, with only a moderate degradation in BER, demonstrating the robustness of the proposed approach for practical MIMO TNM communication systems.
\end{abstract}


\end{frontmatter}

\section{Introduction}
\label{sec:Intro}

Noise Communication (NC) is an emerging communication paradigm for secure and low-power communication, particularly in Internet of Things (IoT) applications \cite{Silva26}. Instead of conventional sinusoidal carrier signals, NC conveys information through noise waveforms whose statistical properties are controlled according to the transmitted information. This approach can potentially eliminate the need for some conventional receiver components, such as local oscillators (LOs) and phase-locked loops (PLLs). The origin of NC can be traced back to the pioneering work of Kish \cite{Kish05}, in which the thermal noise generated by two resistors with low and high resistance values was used to represent binary information through different noise variance levels. Following several extensions of Kish's work, primarily from a physics-oriented perspective \cite{Ming08}-\cite{Kape22}, the work by Basar provided an important development toward establishing noise-based communication as a communication-engineering research topic \cite{Basar23}.

Subsequent studies have explored several directions in noise communication. Fundamental noise modulation and its detection have been investigated in \cite{Basar24,Alshaw24}, while alternative signaling schemes, including on-off, spread-spectrum, and differential noise modulation, have been studied in \cite{Anjos25,Zayy26_Commlett,Anjos26}. Other works have considered generalized noise modulation and higher-dimensional schemes, such as three-dimensional 8-ary noise modulation \cite{Zayy26_WCL,Zayy26_OJCOMS}. In addition, noise modulation has been explored for wireless energy transfer \cite{Yapici25}, and KLJN-based approaches have investigated enhanced-reliability communication, random resistance flipping, and resistor hopping \cite{Salem25,Tasci25,ZayyArxivRH25}.

Several challenges have been investigated in the noise communication framework, including enhancing security \cite{Kish13}-\cite{Melhem19}, increasing signaling dimensionality \cite{Zayy26_Commlett,Zayy26_WCL,Zayy26_OJCOMS,ZayyArxivRH25}, channel estimation \cite{Shen25,Yang26}, and the extension to Multiple-Input Multiple-Output (MIMO) noise communication systems \cite{Yang26}. MIMO communication is a well-established technique in wireless communication systems that exploits multiple antennas to provide spatial diversity and additional spatial degrees of freedom at the receiver \cite{Zheng15}-\cite{Pan24}. In this paper, we focus on the channel estimation problem in MIMO noise communication. In \cite{Shen25}, a modified Maximum Likelihood (ML) channel estimator was proposed together with a Cramér--Rao Bound (CRB) for channel estimation in thermal-noise communication. In \cite{Yang26}, a Majorization-Minimization (MM)-based channel estimation algorithm was developed for MIMO thermal noise communication. More recently, generalized M-ary noise modulation has also been investigated in \cite{Alshaw26}.

Beyond noise-based communication, several related studies have addressed blind detection, noncoherent transmission, MIMO channel estimation, and iterative receiver design. In the context of blind detection, Li et al. \cite{Li21} proposed a clustering-based approach aided by spatially coupled low-density generator matrix codes, where an expectation-maximization (EM) algorithm is employed to estimate the parameters of a system-constrained Gaussian mixture model. For noncoherent multiple-antenna communication, Zheng and Tse \cite{Zheng02} developed a geometric framework based on the Grassmann manifold, establishing fundamental principles for communication without channel state information. In the area of MIMO channel estimation, Li et al. \cite{Li22} investigated a data-aided estimation approach combining clustering and reinforcement learning. From an algorithmic perspective, Fessler and Hero \cite{Fessler94} introduced the space-alternating generalized expectation-maximization (SAGE) algorithm, providing a framework for iterative parameter estimation through generalized maximization steps. Iterative receiver design has also been investigated in coded modulation systems; for example, Yang et al. \cite{Yang24} studied hybrid shaping for bit-interleaved coded modulation with iterative decoding, while Fazeli et al. \cite{Fazeli23} developed a noncoherent receiver for multilevel energy-based modulation using iterative decoding. Although these studies provide relevant foundations for blind detection, noncoherent signaling, channel estimation, and iterative processing, they address different signal models and estimation objectives. In contrast, the present work focuses on MIMO thermal noise modulation and develops a pilot-free GEM-based receiver that jointly estimates the channel covariance factor, mixture probabilities, and unknown transmit variance levels from the received observations.

Blind channel estimation for TNM is considerably more challenging than conventional blind estimation because the receiver must infer the transmitted information from stochastic observations while simultaneously estimating the unknown channel parameters. Moreover, practical TNM systems may employ multiple variance levels to improve spectral efficiency, resulting in an M-ary signaling structure whose statistical model is more involved than that of the binary case. Existing receiver formulations generally assume that the transmit variance levels are perfectly known, an assumption that may be difficult to satisfy in practical implementations because of hardware mismatch, resistor tolerances, temperature variations, and calibration inaccuracies. Consequently, joint blind estimation of the channel matrix and the unknown transmit variance levels remains a challenging problem.

To address these challenges, we develop a GEM-based framework for pilot-free channel Gram matrix estimation in MIMO TNM. The framework estimates the channel Gram matrix $\mathbf{H}\mathbf{H}^H$ and mixture probabilities, and is extended to jointly learn unknown transmit variance levels. The channel-matrix factor and variance-level updates employ gradient-based optimization when closed-form solutions are unavailable. Unlike the pilot-assisted MM-based approach in \cite{Yang26}, the proposed framework does not require pilot symbols or prior knowledge of the transmit variance levels in its variance-learning formulation.

The main contributions of this paper are summarized as follows.

\begin{itemize}

\item A blind receiver is proposed for M-ary thermal noise modulation, eliminating the need for dedicated pilot transmission and the associated pilot overhead. The proposed receiver jointly estimates the channel Gram matrix and the unknown transmit variance levels from the received observations.

\item A Generalized Expectation-Maximization (GEM) framework is developed for joint channel Gram matrix and parameter estimation. Since the maximization step does not admit a closed-form solution, an iterative gradient-ascent optimization procedure is incorporated into the M-step, resulting in a GEM-based receiver.

\item Unlike existing approaches that assume known transmit variance levels, the proposed framework estimates the unknown variance levels directly from the received observations. Although the number of variance levels, $M$, is assumed to be known at the receiver, their actual values are not required to be known a priori.

\item The proposed receiver is applicable to general M-ary TNM signaling and does not require prior knowledge of the transmit variance constellation. Thus, the receiver can learn the variance levels directly from the received data.

\end{itemize}

Simulation results demonstrate that the proposed blind receiver can achieve accurate channel Gram matrix estimation and effective BER performance without pilot symbols. Furthermore, the proposed variance-learning strategy incurs only a moderate performance degradation compared with the ideal case of perfectly known variance levels, demonstrating the potential of the proposed approach for practical MIMO TNM communication systems.

The remainder of this paper is organized as follows. Section \ref{model_system} introduces the system model and formulates the considered MIMO thermal noise modulation framework. Section \ref{proposed_GEM} presents the proposed blind GEM receiver, including the baseline receiver with known variance levels and the proposed variance-learning approach. Section \ref{complexity} analyzes the computational complexity of the proposed algorithm. Section \ref{simulation} presents the simulation results and evaluates the performance of the proposed receiver under different system configurations. Finally, Section \ref{conclusion} concludes the paper.

\section{System Model}
\label{model_system}

Consider a MIMO noise communication system with $N_t$ transmit antennas and $N_r$ receive antennas. Let $k\in\{1,\ldots,N_b\}$ denote the symbol interval index and $l\in\{1,\ldots,L\}$ denote the sample index within a symbol interval. The received signal vector corresponding to the $l$-th sample of the $k$-th symbol interval is modeled as
\begin{equation}
\yb_{k,l}=\Hb\nb_{k,l}+\wb_{k,l},
\end{equation}
where $\Hb\in\mathbb{C}^{N_r\times N_t}$ denotes the unknown channel matrix, $\nb_{k,l}\in\mathbb{C}^{N_t\times 1}$ is the transmitted noise signal vector, and $\wb_{k,l}\in\mathbb{C}^{N_r\times 1}$ represents the additive white Gaussian noise vector.

For the $k$-th symbol interval, the $L$ received samples are collected into the observation matrix
\begin{equation}
\Yb_k=\left[\yb_{k,1},\yb_{k,2},\ldots,\yb_{k,L}\right].
\end{equation}

Similarly, the transmitted signal samples and the receiver noise samples are arranged as
\begin{equation}
\Nb_k=\left[\nb_{k,1},\nb_{k,2},\ldots,\nb_{k,L}\right]
\end{equation}
and
\begin{equation}
\Wb_k=\left[\wb_{k,1},\wb_{k,2},\ldots,\wb_{k,L}\right],
\end{equation}
respectively. Consequently, the received signal model for the $k$-th symbol interval can be compactly expressed as
\begin{equation}
\Yb_k=\Hb\Nb_k+\Wb_k.
\end{equation}

In TNM systems, information is conveyed through the variance levels of the transmitted noise signals. Assuming that the $k$-th symbol interval belongs to the $i$-th variance state, the transmitted noise samples are modeled as
\begin{equation}
\nb_{k,l}\mid z_k=i\sim \mathcal{CN}\left(\mathbf{0},v_i\Ib_{N_t}\right),
\end{equation}
where $v_i$ denotes the variance associated with the $i$-th TNM symbol. Furthermore, the receiver noise is assumed to follow
\begin{equation}
\wb_{k,l}\sim \mathcal{CN}\left(\mathbf{0},\sigma_w^2\Ib_{N_r}\right).
\end{equation}

Throughout this paper, the modulation order $M$ and the receiver-noise variance $\sigma_w^2$ are assumed to be known at the receiver. The modulation order specifies the number of possible TNM variance states, whereas the actual transmit variance levels are unknown and jointly estimated with the channel matrix and mixture probabilities. The receiver noise is assumed to be spatially independent and circularly symmetric complex Gaussian, with a common variance $\sigma_w^2$ across all receive antennas; accordingly, its covariance matrix is given by $\sigma_w^2\mathbf I_{N_r}$. The receiver-noise variance is assumed to be available through prior receiver calibration or estimation during idle listening intervals. Accordingly, the proposed receiver is blind in the sense that it does not require pilot symbols, prior channel knowledge, or prior knowledge of the transmit variance constellation.

Under these assumptions, conditioned on the $i$-th variance state, the received signal vector follows
\begin{equation}
\yb_{k,l}\mid z_k=i \sim \mathcal{CN}\left(\mathbf{0},\Rb_i\right),
\end{equation}
where the covariance matrix is given by
\begin{equation}
\Rb_i=\mathbb{E}\left[\yb_{k,l}\yb_{k,l}^{H}\mid z_k=i\right]=v_i\Hb\Hb^{H}+\sigma_w^2\Ib_{N_r}.
\end{equation}

The hidden variance state is defined as
\begin{equation}
z_k=i,
\end{equation}
where $i\in\{1,\ldots,M\}$ denotes the variance level used during the $k$-th symbol interval. The prior probability of the $i$-th state is given by
\begin{equation}
P(z_k=i)=\pi_i,\qquad\sum_{i=1}^{M}\pi_i=1.
\end{equation}

For the subsequent likelihood formulation, the received samples within the $k$-th symbol interval are summarized by the scatter matrix
\begin{equation}
\Sbb_k=\Yb_k\Yb_k^{H}.
\end{equation}

\section{Proposed Receiver}
\label{proposed_GEM}

This section presents the proposed blind receiver for M-ary thermal noise modulation. The receiver is developed within the Generalized Expectation--Maximization (GEM) framework and iteratively estimates the unknown channel parameters and latent variance states from the received observations. To facilitate the presentation, the receiver is introduced in two stages. First, a baseline receiver is derived under the assumption that the transmit variance levels are perfectly known. Subsequently, this formulation is extended to the practical scenario in which the variance levels are also unknown and jointly estimated with the channel parameters.

\subsection{GEM Receiver with Known Variance Levels}

In this section, a blind receiver is developed under the assumption that the variance levels of the TNM constellation are perfectly known at the receiver. The proposed receiver follows the Expectation--Maximization (EM) framework \cite{Jeff83,Yang23} by iteratively estimating the hidden variance states in the E-step and updating the unknown parameters in the M-step through maximization of the expected complete-data log-likelihood. Since the channel parameters appear nonlinearly in the covariance matrix, the M-step does not admit a closed-form solution. Therefore, an iterative gradient-ascent procedure is employed within the M-step, resulting in a Generalized Expectation--Maximization (GEM) algorithm.

For the known-variance case, the unknown parameter set is defined as
\begin{equation}
\Theta=\left\{\Hb,\pi_1,\ldots,\pi_M\right\},
\end{equation}
where the variance levels $\{v_i\}_{i=1}^{M}$ are assumed to be known.

Conditioned on the variance state $z_k=i$, the likelihood of the received samples in the $k$-th symbol interval is
\begin{equation}
p(\Yb_k|z_k=i,\Hb)=\prod_{l=1}^{L}\mathcal{CN}\left(\yb_{k,l};\mathbf{0},\Rb_i\right),
\end{equation}
where
\begin{equation}
\Rb_i=v_i\Hb\Hb^{H}+\sigma_w^2\Ib_{N_r}.
\end{equation}

Since the variance state is hidden, the corresponding incomplete-data likelihood is
\begin{equation}
p(\Yb_k|\Hb,\boldsymbol{\pi})=\sum_{i=1}^{M}\pi_i p(\Yb_k|z_k=i,\Hb),
\end{equation}
where
\begin{equation}
\boldsymbol{\pi}=[\pi_1,\ldots,\pi_M]^T.
\end{equation}

Using the definition
\begin{equation}
\Sbb_k=\Yb_k\Yb_k^{H},
\end{equation}
the conditional log-likelihood can be written, up to an additive constant, as
\begin{equation}
\log p(\Yb_k|z_k=i,\Hb)=-L\log|\Rb_i|-\operatorname{tr}\left(\Rb_i^{-1}\Sbb_k\right)+\mathrm{const}.
\end{equation}

If the hidden states were known, the complete-data likelihood would be
\begin{equation}
p(\Yb,\Zb|\Hb,\boldsymbol{\pi})=\prod_{k=1}^{N_b}\prod_{i=1}^{M}\left[\pi_ip(\Yb_k|z_k=i,\Hb)\right]^{\mathbb{I}(z_k=i)},
\end{equation}
and the corresponding complete-data log-likelihood is
\begin{equation}
\begin{aligned}
\log p(\Yb,\Zb|\Hb,\boldsymbol{\pi})=\sum_{k=1}^{N_b}\sum_{i=1}^{M}\mathbb{I}(z_k=i)\Big[\log\pi_i+\log p(\Yb_k|z_k=i,\Hb)\Big],
\end{aligned}
\end{equation}
where $\mathbb{I}(\cdot)$ denotes the indicator function.

\begin{enumerate}[\label={}]

\item \textbf{E-Step:}

The posterior responsibility of the $i$-th variance state for the $k$-th observation block is defined as
\begin{equation}
r_{ki}=P(z_k=i|\Yb_k,\Theta^{\mathrm{old}}).
\end{equation}

Using Bayes' rule, the responsibility is obtained as
\begin{equation}
r_{ki}=\frac{\pi_i|\Rb_i|^{-L}\exp\!\left[-\operatorname{tr}\left(\Rb_i^{-1}\Sbb_k\right)\right]}{\displaystyle\sum_{j=1}^{M}\pi_j|\Rb_j|^{-L}\exp\!\left[ -\operatorname{tr} \left(\Rb_j^{-1}\Sbb_k\right)\right]}.
\end{equation}

The EM auxiliary function is given by
\begin{equation}
Q(\Theta|\Theta^{\mathrm{old}})=\mathbb{E}_{\Zb|\Yb,\Theta^{\mathrm{old}}}\left[\log p(\Yb,\Zb|\Theta)\right].
\end{equation}

Replacing the indicator function by its posterior expectation yields
\begin{equation}
Q(\Theta|\Theta^{\mathrm{old}})=\sum_{k=1}^{N_b}\sum_{i=1}^{M}r_{ki}\left[\log p(\Yb_k|z_k=i,\Hb)+\log\pi_i\right].
\end{equation}

Substituting the Gaussian likelihood gives
\begin{align}
Q(\Theta|\Theta^{\mathrm{old}})=&-\sum_{k=1}^{N_b}\sum_{i=1}^{M}r_{ki}\left[L\log|\Rb_i|+\operatorname{tr}\left(\Rb_i^{-1}\Sbb_k\right)\right]\nonumber\\
&+\sum_{k=1}^{N_b}\sum_{i=1}^{M}r_{ki}\log\pi_i+\mathrm{const}.
\end{align}

\item \textbf{M-Step:}

In the M-step, the auxiliary function is maximized with respect to the unknown channel matrix and mixture probabilities. Since the variance levels are assumed to be known, they are not updated in this stage.

The channel matrix is updated according to
\begin{equation}
\Hb^{\mathrm{new}}=\Hb^{\mathrm{old}}+\mu_H\nabla_{\Hb^{*}}Q,
\end{equation}
where $\mu_H>0$ denotes the step size. The channel-update step size $\mu_H$ is kept fixed throughout each run and is selected empirically according to the simulation conditions. No line-search procedure or automatic adaptive step-size rule is employed. The gradient of the auxiliary function with respect to $\Hb^{*}$ is given by
\begin{equation}
\nabla_{\Hb^{*}}Q=\sum_{k=1}^{N_b}\sum_{i=1}^{M}r_{ki}v_i\left[\Rb_i^{-1}\Sbb_k\Rb_i^{-1}\Hb-L\Rb_i^{-1}\Hb\right].
\end{equation}

To reduce the computational cost associated with repeated matrix inversion, the inverse of $\Rb_i$ can be evaluated using the Woodbury identity as
\begin{equation}
\Rb_i^{-1}=\frac{1}{\sigma_w^2}\left[\Ib_{N_r}-\Hb\left(\frac{\sigma_w^2}{v_i}\Ib_{N_t}+\Hb^{H}\Hb\right)^{-1}\Hb^{H}\right].
\end{equation}

The mixture probabilities are updated in closed form as
\begin{equation}
\pi_i^{\mathrm{new}}=\frac{1}{N_b}\sum_{k=1}^{N_b}r_{ki},\qquad i=1,\ldots,M.
\end{equation}

The E- and M-steps are repeated until the convergence criterion is satisfied. This known-variance formulation provides the baseline for the proposed variance-level learning scheme described next.

\end{enumerate}

\subsection{GEM Receiver with Variance-Level Learning}

The known-variance formulation presented above assumes that the transmit variance constellation is perfectly known at the receiver. In practice, however, the actual variance levels may be unknown because of hardware imperfections, calibration errors, or variations in the transmission conditions. Therefore, the proposed receiver is extended to jointly estimate the channel matrix, mixture probabilities, and variance levels.

The unknown parameter set is defined as
\begin{equation}
\Theta=\left\{\Hb,\pi_1,\ldots,\pi_M,v_1,\ldots,v_M\right\}.
\end{equation}

In this case, the E-step retains the same probabilistic form as in the known-variance formulation, but the covariance matrices $\Rb_i$ are evaluated using the current estimates of both $\Hb$ and $v_i$. Thus,
\begin{equation}
r_{ki}=P(z_k=i|\Yb_k,\Theta^{\mathrm{old}})
\end{equation}
is computed according to
\begin{equation}
r_{ki}=\frac{\pi_i|\Rb_i|^{-L}\exp\!\left[-\operatorname{tr}\left(\Rb_i^{-1}\Sbb_k\right)\right]}{\displaystyle\sum_{j=1}^{M}\pi_j|\Rb_j|^{-L} \exp\!\left[-\operatorname{tr}\left(\Rb_j^{-1}\Sbb_k\right)\right]}.
\end{equation}

The corresponding auxiliary function is
\begin{align}
Q(\Theta|\Theta^{\mathrm{old}})=&-\sum_{k=1}^{N_b}\sum_{i=1}^{M}r_{ki}\left[L\log|\Rb_i|+\operatorname{tr}\left(\Rb_i^{-1}\Sbb_k\right)\right]\nonumber\\
&+\sum_{k=1}^{N_b}\sum_{i=1}^{M}r_{ki}\log\pi_i+\mathrm{const},
\end{align}
where
\begin{equation}
\Rb_i=v_i\Hb\Hb^{H}+\sigma_w^2\Ib_{N_r}.
\end{equation}

The channel matrix is updated using the same gradient-ascent procedure as in the known-variance case. In addition, the variance levels are updated by gradient ascent. The derivative of the auxiliary function with respect to $v_i$ is given by
\begin{equation}
\frac{\partial Q}{\partial v_i}=-\sum_{k=1}^{N_b}r_{ki}\left[L\operatorname{tr}\left(\Rb_i^{-1}\Hb\Hb^{H}\right)- \operatorname{tr}\left(\Rb_i^{-1}\Hb\Hb^{H}\Rb_i^{-1}\Sbb_k\right)\right].
\end{equation}

Accordingly, the variance levels are updated according to
\begin{equation}
v_i^{\mathrm{new}}=v_i^{\mathrm{old}}+\mu_v\frac{\partial Q}{\partial v_i},
\end{equation}
where $\mu_v>0$ denotes the variance-level step size. The selection of the step size $\mu_v$ is discussed in Section~\ref{convergence}, along with the selection of $\mu_H$ and their effects on the convergence behavior. Since the variance levels must be positive, the updated values are projected onto the positive orthant:
\begin{equation}
v_i^{\mathrm{new}} \leftarrow \max \left(v_i^{\mathrm{new}},\epsilon \right),\qquad \epsilon>0.
\end{equation}

The mixture probabilities are updated as
\begin{equation}
\pi_i^{\mathrm{new}}=\frac{1}{N_b}\sum_{k=1}^{N_b}r_{ki}.
\end{equation}

An inherent scale ambiguity exists between the channel matrix and the variance levels. Specifically, for any $c>0$,
\begin{equation}
\Hb\rightarrow\sqrt{c}\Hb, \qquad v_i\rightarrow\frac{v_i}{c}
\end{equation}
leaves the covariance matrix $\Rb_i$ unchanged. Therefore, the absolute scale of $\Hb$ and $\{v_i\}$ cannot be identified independently from the received observations. To fix this ambiguity, the estimated variance levels are normalized after each iteration. Specifically, let
\begin{equation}
c=\min_i v_i.
\end{equation}
Then,
\begin{equation}
v_i\leftarrow\frac{v_i}{c},\qquad \Hb\leftarrow\sqrt{c}\Hb.
\end{equation}
This normalization fixes the arbitrary scale by setting the smallest variance level to unity while preserving the covariance matrices $\Rb_i$.

Since the ordering of the variance states is not identifiable from the likelihood, the estimated variance levels are sorted in ascending order at the end of each GEM iteration. The same permutation is applied to the corresponding mixture probabilities and posterior responsibilities to maintain consistent state labeling. A predefined Gray mapping between the ordered variance states and binary labels is assumed to be shared by the transmitter and receiver. Accordingly, the receiver associates the detected variance states with their corresponding binary labels based on this mapping, thereby resolving the state-label ambiguity without requiring pilot symbols.

It is worth noting that the above normalization resolves only the scale ambiguity between the channel matrix and the variance levels. However, the channel matrix itself is not uniquely identifiable under the assumed Gaussian signaling model. Specifically, for any unitary matrix $\mathbf{U}\in\mathbb{C}^{N_t\times N_t}$, replacing $\mathbf{H}$ with $\mathbf{H}\mathbf{U}$ leaves $\mathbf{H}\mathbf{H}^{H}$, and hence all received covariance matrices, unchanged. Therefore, the proposed GEM algorithm estimates the channel covariance matrix $\mathbf{H}\mathbf{H}^{H}$ through a channel-matrix factorization, rather than uniquely recovering the physical channel matrix. The estimated \(\mathbf{H}\) should thus be interpreted as one possible representative of the equivalence class of channel matrices that yield the same received covariance.

More specifically, when the transmit variance levels and the receiver-noise variance are known, the matrix $\mathbf G=\mathbf H\mathbf H^H$ can be uniquely determined from the component covariance matrices as
\begin{equation}
\mathbf G=\frac{\mathbf R_i-\sigma_w^2\mathbf I_{N_r}}{v_i},
\end{equation}
for any $v_i>0$. However, when the transmit variance levels are also unknown, only the products $v_i\mathbf G$ are directly identifiable from the component covariance matrices. In this case, the scale normalization described above selects a unique normalized representation by setting the smallest variance level to unity. Thus, the normalized matrix $\mathbf G$ and the normalized variance levels are identifiable, subject to the assumed Gaussian mixture model and the usual permutation ambiguity of its components.

The resulting GEM receiver jointly estimates a factorization of the channel Gram matrix, the mixture probabilities, and the normalized transmit variance levels, without requiring prior knowledge of the transmit variance constellation or pilot symbols.

The overall procedure of the proposed GEM-based blind receiver is summarized in Algorithm~\ref{alg:GEM}. The algorithm iteratively performs the E-step and M-step to jointly estimate the channel matrix, variance levels, and mixture probabilities. The positivity constraint, scale normalization, and state-label ordering are also incorporated into the iterative procedure.

\begin{algorithm}[tb]
\caption{Proposed GEM-Based Blind Channel and Variance-Level Estimation}
\label{alg:GEM}
\begin{algorithmic}[1]

\State \textbf{Input:}
Received observations $\{\Yb_k\}_{k=1}^{K}$,
known modulation order $M$,
known receiver-noise variance $\sigma_w^2$,
maximum number of iterations $T_{\max}=3000$,
step sizes $\mu_H$ and $\mu_v$.
and positivity parameter $\epsilon=10^{-6}$.

\State \textbf{Initialize:} $\Hb^{(0)}$, $\{v_i^{(0)}\}_{i=1}^{M}$, and $\{\pi_i^{(0)}\}_{i=1}^{M}$.

\For{$t=0,1,\ldots,T_{\max}-1$}

    \State \textbf{E-step:} Compute the posterior responsibilities
    \begin{equation*}
    r_{ki}^{(t)}=P(z_k=i|\Yb_k,\Theta^{(t)})
    \end{equation*}
    for $k=1,\ldots,N_b$ and $i=1,\ldots,M$.

    \State Compute
    \begin{equation*}
    \Rb_i^{(t)}=v_i^{(t)}\Hb^{(t)}\Hb^{(t)H}+\sigma_w^2\Ib_{N_r}.
    \end{equation*}

    \State \textbf{M-step:} Update the channel matrix using gradient ascent:
    \begin{equation*}
    \Hb^{(t+1)}=\Hb^{(t)}+\mu_H\nabla_{\Hb^*}Q.
    \end{equation*}

    \State Update each variance level using gradient ascent:
    \begin{equation*}
    v_i^{(t+1)}=v_i^{(t)}+\mu_v
    \frac{\partial Q}{\partial v_i}.
    \end{equation*}

    \State Project the variance levels onto the positive orthant:
    \begin{equation*}
    v_i^{(t+1)}\leftarrow \max\left(v_i^{(t+1)},\epsilon\right).
    \end{equation*}

    \State Update the mixture probabilities:
    \begin{equation*}
    \pi_i^{(t+1)}=\frac{1}{N_b}\sum_{k=1}^{N_b}r_{ki}^{(t)}.
    \end{equation*}

    \State Resolve the scale ambiguity by setting
    \begin{equation*}
    c=\min_i v_i^{(t+1)},
    \end{equation*}
    followed by
    \begin{equation*}
    v_i^{(t+1)}\leftarrow\frac{v_i^{(t+1)}}{c},\qquad\Hb^{(t+1)}\leftarrow\sqrt{c}\Hb^{(t+1)}.
    \end{equation*}

    \State Sort $\{v_i^{(t+1)}\}_{i=1}^{M}$ in ascending order and apply the same permutation to $\{\pi_i^{(t+1)}\}$ and $\{r_{ki}^{(t)}\}_{k=1}^{N_b}$.

\EndFor

\State \textbf{Output:} $\hat{\Hb}$, $\hat{\mathbf{v}}$, and $\hat{\boldsymbol{\pi}}$.

\end{algorithmic}
\end{algorithm}

\subsection{Bit Detection}

After the GEM algorithm converges, the final estimates $\hat{\Hb}$, $\hat{\mathbf v}$, and $\hat{\boldsymbol\pi}$ are employed for symbol detection. For each received observation block $\Yb_k$, the posterior probabilities are computed as
\begin{equation}
r_{ki}=P(z_k=i|\Yb_k,\hat{\Theta}),
\end{equation}
where
\begin{equation}
\hat{\Theta}={\hat{\Hb},\hat{\mathbf v},\hat{\boldsymbol\pi}}.
\end{equation}

The transmitted variance state is detected according to the maximum a posteriori (MAP) rule
\begin{equation}
\hat z_k=\arg\max_i r_{ki}.
\end{equation}

Each detected variance state is subsequently mapped to its corresponding information label according to the adopted M-ary TNM constellation \cite{Alshaw26}. For an $M$-ary TNM system with $M=2^b$, each detected symbol carries $b=\log_2 M$ bits. Finally, the bit-error rate (BER) is computed by comparing the detected bit sequence with the transmitted bit sequence over all observation intervals.

\section{Theoretical Analysis}
\label{complexity}

In this section, the theoretical properties of the proposed GEM-based receiver are investigated. First, the convergence behavior of the proposed iterative estimation algorithm is analyzed. Then, the computational complexity of the proposed receiver is derived and discussed with respect to the system dimensions and the number of variance states.

\subsection{Convergence Analysis}
\label{convergence}
The proposed receiver belongs to the class of Generalized Expectation-Maximization (GEM) algorithms because the M-step updates the unknown parameters through iterative gradient-ascent optimization rather than exact maximization of the EM auxiliary function. 

Let the unknown parameter set be defined as
\begin{equation}
\Theta=\{\Hb,\vb,\boldsymbol{\pi}\},
\end{equation}
where
\begin{equation}
\vb=[v_1,\ldots,v_M]^T.
\end{equation}

The observed-data log-likelihood is given by
\begin{equation}
\ell(\Theta)=\log p(\Yb;\Theta),
\end{equation}
where the latent variables correspond to the unknown variance states
\begin{equation}
\Zb={z_1,\ldots,z_{N_b}}.
\end{equation}

For the current parameter estimate $\Theta^{(t)}$, the EM auxiliary function is defined as
\begin{equation}
Q(\Theta|\Theta^{(t)})=\mathbb{E}_{\Zb|\Yb,\Theta^{(t)}}\left[\log p(\Yb,\Zb;\Theta)
\right].
\end{equation}

Using the posterior probabilities
\begin{equation}
r_{ki}=P(z_k=i|\Yb_k,\Theta^{(t)}),
\end{equation}
the auxiliary function can be written as
\begin{equation}
Q(\Theta|\Theta^{(t)})=\sum_{k=1}^{N_b}\sum_{i=1}^{M}r_{ki}\left[\log p(\Yb_k|z_k=i;\Theta)+\log\pi_i\right].
\end{equation}

The covariance matrix corresponding to the $i$-th variance state is
\begin{equation}
\Rb_i=v_i\Hb\Hb^H+\sigma_w^2\Ib_{N_r}.
\end{equation}

Since $v_i>0$ and $\sigma_w^2>0$, for any nonzero $\xb$,
\begin{equation}
\xb^H\Rb_i\xb=v_i|\Hb^H\xb|^2+\sigma_w^2|\xb|^2.
\end{equation}
Hence,
\begin{equation}
\Rb_i\succ0,
\end{equation}
and, more specifically,
\begin{equation}
\Rb_i\succeq\sigma_w^2\Ib_{N_r}.
\end{equation}

Therefore, $\Rb_i$ is nonsingular for all admissible parameter values, and the terms $\log|\Rb_i|$ and $\operatorname{tr}(\Rb_i^{-1}\Sbb_k)$ are well defined throughout the iterations.

For the convergence analysis of the gradient-based updates, assume that the auxiliary function is continuously differentiable and that its partial gradients with respect to the channel matrix and the variance levels are Lipschitz continuous in the equivalent real-valued parameter space. Specifically, let $L_H>0$ and $L_v>0$ denote the corresponding Lipschitz constants for the channel and variance updates, respectively. The complex-valued channel matrix is understood through its equivalent real-valued representation.

For a fixed set of posterior responsibilities, the standard ascent lemma gives the following lower bound for a channel update:
\begin{equation}
Q(\mathbf H+\Delta_H,\mathbf v) \geq Q(\mathbf H,\mathbf v) +\langle\nabla_{\mathbf H}Q,\Delta_H\rangle -\frac{L_H}{2}\|\Delta_H\|_F^2.
\end{equation}

Choosing the gradient-ascent direction as
\begin{equation}
\Delta_H=\mu_H\nabla_{\mathbf H}Q
\end{equation}

yields
\begin{equation}
Q(\mathbf H^{\mathrm{new}},\mathbf v) -Q(\mathbf H^{\mathrm{old}},\mathbf v) \geq \mu_H\left(1-\frac{L_H\mu_H}{2}\right) \|\nabla_{\mathbf H}Q\|_F^2.
\end{equation}

Similarly, for the variance-level update,
\begin{equation}
Q(\mathbf H,\mathbf v+\Delta_v) \geq Q(\mathbf H,\mathbf v) +\nabla_{\mathbf v}Q^T\Delta_v -\frac{L_v}{2}\|\Delta_v\|_2^2.
\end{equation}

By choosing $\Delta_v=\mu_v\nabla_{\mathbf v}Q$, we obtain
\begin{equation}
Q(\mathbf H,\mathbf v^{\mathrm{new}}) -Q(\mathbf H,\mathbf v^{\mathrm{old}}) \geq \mu_v\left(1-\frac{L_v\mu_v}{2}\right) \|\nabla_{\mathbf v}Q\|_2^2.
\end{equation}

Consequently, the sufficient conditions
\begin{equation}
0<\mu_H<\frac{2}{L_H}, \qquad 0<\mu_v<\frac{2}{L_v}
\end{equation}
ensure non-decreasing auxiliary-function values for the corresponding unconstrained gradient-ascent updates. The exact update of the mixture probabilities also maximizes the auxiliary function with respect to $\{\pi_i\}$, subject to the probability-simplex constraints. Therefore, if the successive parameter updates satisfy the corresponding ascent conditions, the resulting GEM update satisfies
\begin{equation}
Q(\Theta^{(t+1)}|\Theta^{(t)})
\geq Q(\Theta^{(t)}|\Theta^{(t)}).
\end{equation}

Moreover, since $\mathbf R_i\succeq\sigma_w^2\mathbf I_{N_r}$, with $\sigma_w^2>0$, the Gaussian likelihood of each observation block is bounded above by a finite value. Thus, the observed-data log-likelihood is bounded above over the admissible parameter space. Under the standard GEM regularity conditions, if the auxiliary-function ascent condition is satisfied at every outer iteration, the observed-data log-likelihood sequence is non-decreasing and converges to a finite limit.

In the numerical implementation, a single gradient-ascent update is performed for each parameter block at every outer iteration. The step sizes $\mu_H$ and $\mu_v$ are selected empirically through preliminary numerical experiments and kept fixed during each simulation run. No line-search procedure is employed, and the Lipschitz constants are not explicitly evaluated. Therefore, although the derived bounds provide sufficient conditions for monotonic ascent, their satisfaction by the empirically selected step sizes is not explicitly verified. Accordingly, the theoretical convergence result is conditional on these sufficient ascent conditions, while the convergence behavior of the practical implementation is examined numerically.

\subsection{Complexity Analysis}

\noindent\textbf{Computational Complexity:}
The computational complexity of the proposed GEM receiver is analyzed in terms of the number of variance states $M$, the number of observation blocks $N_b$, the number of received samples per block $L$, the MIMO dimensions $N_t$ and $N_r$, and the number of GEM iterations $I$.

Before the iterative GEM procedure, the sample covariance-related matrices are computed from the received observations as
$\mathbf S_k=\mathbf Y_k\mathbf Y_k^H$, where $\mathbf Y_k\in\mathbb C^{N_r\times L}$ contains the $L$ received samples in the $k$-th observation block. The computation of these matrices requires $\mathcal O(N_r^2L)$ operations per block, resulting in a total preprocessing complexity of $\mathcal O(N_b N_r^2L)$.

Since the matrices $\mathbf S_k$ are computed only once and reused throughout the iterative GEM procedure, this cost is not incurred in every GEM iteration.

During the E-step, the posterior responsibilities are computed for each observation block and each variance state. Owing to the Woodbury matrix identity, direct inversion of the covariance matrix $\Rb_i\in\mathbb{C}^{N_r\times N_r}$ is avoided. Instead, only the inversion of the smaller matrix $\left(\frac{\sigma_w^2}{v_i}\Ib_{N_t}+\Hb^{H}\Hb\right)$ of dimension $N_t\times N_t$ is required. Thus, the computational cost associated with matrix inversion is reduced, particularly when $N_r\gg N_t$.

The inversion of the reduced $N_t\times N_t$ matrix is performed once per variance state in each GEM iteration, requiring $\mathcal O(MN_t^3)$ operations.

The dominant computational cost of the E-step is associated with evaluating the quadratic trace terms $\operatorname{tr}\left(\Rb_i^{-1}\Sbb_k\right)$ for all $(k,i)$ pairs. When the covariance inverses are explicitly formed, this operation requires $\mathcal{O}(N_r^2)$ computations per $(k,i)$ pair, resulting in an overall E-step complexity of $\mathcal{O}(MN_bN_r^2)$.

In the M-step, the most computationally demanding operation is the evaluation of the channel gradient
\begin{equation}
\nabla_{\Hb^*}Q=\sum_{k=1}^{N_b}\sum_{i=1}^{M}r_{ki}v_i\left(\Rb_i^{-1}\Sbb_k\Rb_i^{-1}\Hb-L\Rb_i^{-1}\Hb\right),
\end{equation}
whose dominant matrix multiplications require approximately $\mathcal{O}(MN_bN_r^2N_t)$ operations.

The updates of the variance levels and mixture probabilities involve trace evaluations and weighted summations. Their computational costs are of order
$\mathcal{O}(MN_bN_r^2)$ and $\mathcal{O}(MN_b)$,
respectively.

Including the preprocessing cost, the overall computational complexity of the proposed receiver after $I$ GEM iterations is given by
$\mathcal{O}\left(N_bN_r^2L+IMN_bN_r^2N_t\right)$. For square MIMO systems with $N_t=N_r=N$, this expression reduces to $\mathcal{O}\left(N_b N^2L+IMN_bN^3\right)$.

\begin{table}[t]
\centering
\caption{Computational complexity comparison of channel
estimation algorithms.}
\label{tab:channel_estimation_complexity}
\renewcommand{\arraystretch}{1.5}
\setlength{\tabcolsep}{5pt}
\begin{tabular}{|c|c|c|}
\hline
\textbf{Method} &
\textbf{Estimation Type} &
\textbf{Computational Complexity} \\
\hline
MM \cite{Yang26} &
Pilot-assisted &
\makecell[l]{
$\mathcal{O}\big(L_pN_r^2N_s+
I_{\mathrm{power}}N_r^2N_t$\\[4pt]
$\qquad+
I_{\mathrm{MM}}L_p
(N_r^2N_t+N_rN_t^2+N_t^3)\big)$
}
\\
\hline
Proposed GEM &
Blind (Pilot-free) &
$\mathcal{O}\big(
N_bN_r^2L+
I_{\mathrm{GEM}}M_vN_bN_r^2N_t
\big)$
\\
\hline
\end{tabular}
\end{table}

Table~\ref{tab:channel_estimation_complexity} compares the computational complexity of the proposed GEM-based method with the MM-based channel estimation algorithm in \cite{Yang26}. Unlike the pilot-assisted method in \cite{Yang26}, the proposed approach performs pilot-free channel estimation while jointly estimating the channel matrix and the unknown variance levels.

\noindent\textbf{Memory Complexity:}
Regarding memory requirements, the proposed implementation stores the sample covariance matrices $\mathbf S_k$ for all $N_b$ observation blocks, requiring $\mathcal O(N_bN_r^2)$ memory locations. The posterior responsibility matrix, whose dimensions are $N_b\times M$, requires $\mathcal O(N_bM)$ memory locations. In addition, storing the channel matrix and the state-dependent reduced matrices requires $\mathcal O(N_rN_t)$ and $\mathcal O(MN_t^2)$ memory locations, respectively. Therefore, the overall memory complexity of the implemented receiver is given by $\mathcal O(N_bN_r^2+N_bM+N_rN_t+MN_t^2)$. This expression corresponds to an implementation that stores the sufficient statistics and posterior responsibilities for all observation blocks. The memory requirement can be reduced by processing the observation blocks sequentially.

It is worth emphasizing that the use of the Woodbury identity avoids repeated inversion of full $N_r\times N_r$ covariance matrices during the iterative optimization. This substantially reduces the computational burden when the number of receive antennas is larger than the number of transmit antennas.

\section{Simulation Results}\label{simulation}
This section presents the simulation results of the proposed GEM-based framework for joint channel estimation and variance-level learning in MIMO noise communication systems. The performance of the proposed algorithm is evaluated under different SNR values and MIMO configurations and is compared with the ideal case in which the variance levels are perfectly known. Unless otherwise specified, all simulations are performed over Rayleigh fading channels with additive white Gaussian noise using $N_b$ received signal blocks, each consisting of $L$ noise samples. In all experiments, four distinct variance states with variance levels $\{1,2,4,8\}$ are employed, representing a 2-bit variance-state signaling scheme where each transmitted block belongs to one of four possible variance classes. The effectiveness of the proposed approach is assessed in terms of the classification error rate (CER), bit error rate (BER), the normalized mean square error (NMSE) of the estimated channel Gram matrix, and the NMSE of the estimated variance levels. The CER is defined as the ratio of incorrectly classified variance states to the total number of transmitted signal blocks:
\begin{equation}
CER=\frac{1}{N_b}\sum_{k=1}^{N_b}\mathbf{1}(\hat{z}_k\neq z_k),
\end{equation}
where $z_k$ and $\hat{z}_k$ denote the true and estimated variance-state indices, respectively. The NMSE of the channel Gram matrix and the variance levels are defined as
\begin{equation}
\mathrm{NMSE}_{\mathbf G}=\frac{\|\widehat{\mathbf H}\widehat{\mathbf H}^H-\mathbf H\mathbf H^H\|_F^2}
{\|\mathbf H\mathbf H^H\|_F^2},
\quad
\mathrm{NMSE}_{v}=
\frac{\|\widehat{\mathbf v}-\mathbf v\|_2^2}
{\|\mathbf v\|_2^2}.
\end{equation}

The scale ambiguity is handled through the normalization procedure described in Section~3.2, and the true and estimated parameters are compared under the same normalization convention. Therefore, no additional scale alignment is performed before NMSE evaluation.

For clarity and reproducibility, the main simulation parameters and configurations used in the numerical experiments are summarized in Table~\ref{tab:simulation_parameters}. Some parameters, such as the MIMO dimensions, number of observation blocks, and block length, vary across experiments according to the performance metric under investigation. The corresponding settings are specified in the table, while the algorithmic step sizes are selected separately for each simulation scenario to ensure stable numerical updates.

\begin{table}[!t]
\centering
\caption{Main simulation parameters and configurations.}
\label{tab:simulation_parameters}
\renewcommand{\arraystretch}{1.2}
\begin{tabular}{|l| l|}
\hline
\textbf{Parameter} & \textbf{Setting} \\
\hline
Channel model & Rayleigh fading \\
\hline
Noise model & AWGN \\
\hline
Number of variance states ($M$) & 4 \\
\hline
Variance levels ($\mathbf{v}$) & $\{1,2,4,8\}$ \\
\hline
MIMO configurations ($N_t\times N_r$)
& $2\times2,\ 4\times4,\ 4\times 8$ \\
\hline
Number of observation blocks ($N_b$)
& 3000, 1500, 512, or variable \\
\hline
Samples per block ($L$)
& 40 or variable \\
\hline
Maximum GEM iterations & 3000 \\
\hline
Initialization & Same initialization procedure \\
\hline
Step sizes ($\mu_H,\mu_v$)
& Selected separately for each experiment \\
\hline
Performance metrics & CER, BER, NMSE \\
\hline
\end{tabular}
\end{table}

The performance of the proposed receiver is compared with the benchmark methods presented in \cite{Yang26}. Since the reference reports the BER performance in terms of $E_b/N_0$, the same metric is adopted in this paper to enable a fair comparison. Although the proposed simulations are performed by varying the normalized SNR through the receiver noise variance, the corresponding $E_b/N_0$ values are obtained according to the following derivation. In the proposed TNM system, one transmitted symbol consists of $L$ independent noise samples, while each symbol conveys $\log_2(M)$ information bits. Let $E_s$ denote the average received energy corresponding to one TNM symbol. The average energy per information bit is therefore

\begin{equation}
E_b=\frac{E_s}{\log_2(M)}.
\label{eq:Eb}
\end{equation}

Throughout this paper, the receiver noise variance is selected according to

\begin{equation}
\sigma_w^2=\frac{E_s}{10^{\mathrm{SNR}/10}},
\label{eq:sigma}
\end{equation}
which implies that the simulation SNR is defined as

\begin{equation}
\mathrm{SNR}=\frac{E_s}{N_0},
\label{eq:snrdef}
\end{equation}
where $N_0$ denotes the one-sided noise power spectral density. Substituting (\ref{eq:Eb}) into (\ref{eq:snrdef}) yields

\begin{equation}
\frac{E_b}{N_0}=\frac{E_s/N_0}{\log_2(M)}=\frac{\mathrm{SNR}}{\log_2(M)}.
\end{equation}

Expressed in decibels, the relationship becomes

\begin{equation}
\left(\frac{E_b}{N_0}\right)_{\rm dB}=\mathrm{SNR}_{\rm dB}-10\log_{10}\!\left(\log_2M\right).
\label{eq:EbN0}
\end{equation}

\begin{figure}[!t]
\centering
{\includegraphics[width=.9\linewidth]{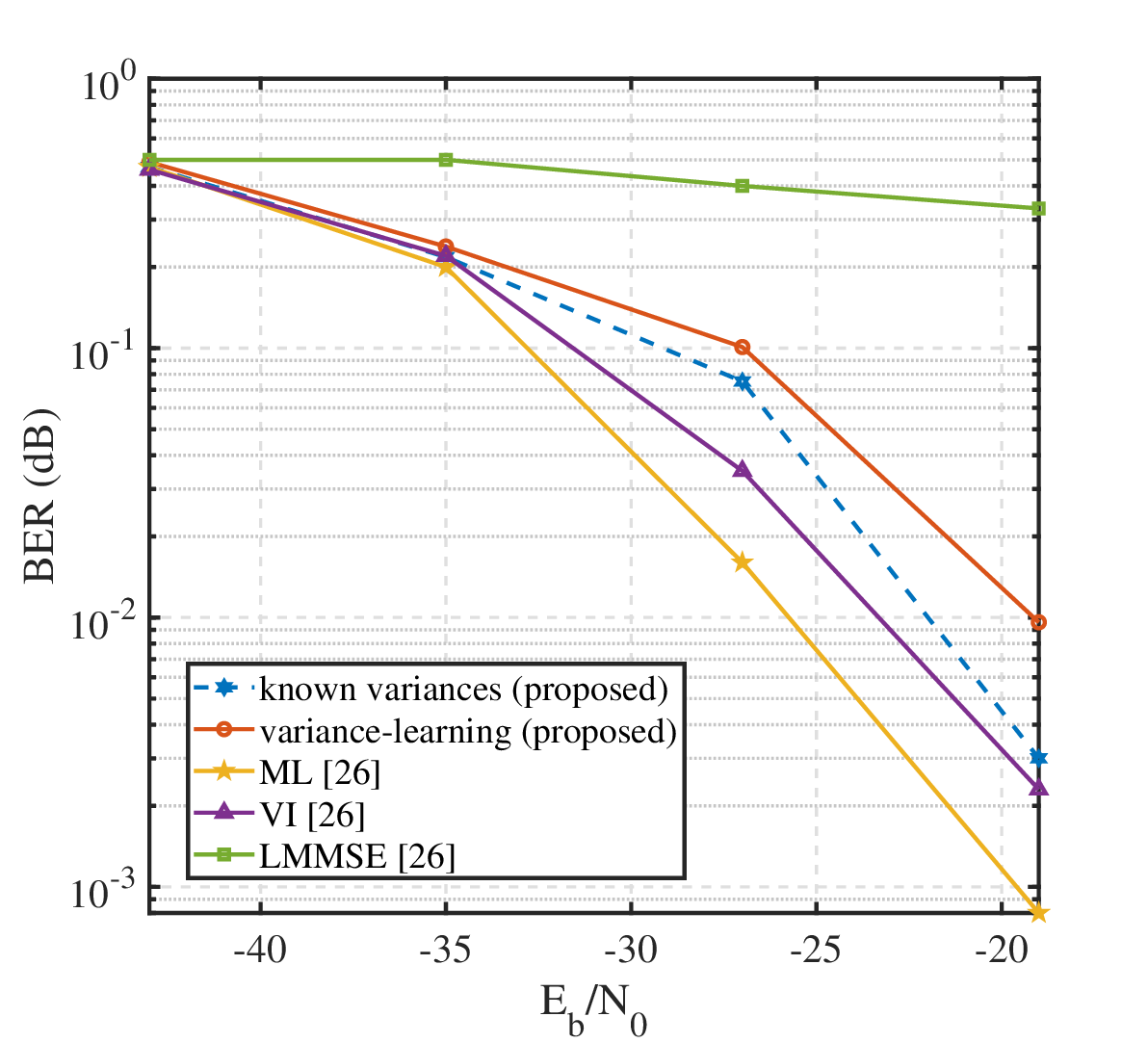}}%
\caption{BER performance of the proposed blind GEM receiver compared with the pilot-aided detectors reported in \cite{Yang26} for $N_b = 512$.}
\label{compare_MM}
\end{figure}

Consequently, all BER results reported in this paper are generated by varying the simulation SNR according to (\ref{eq:sigma}), while the
corresponding $E_b/N_0$ values shown in the Fig. \ref{compare_MM} are obtained from (\ref{eq:EbN0}) to facilitate comparison with \cite{Yang26}.

\begin{figure}[!t]
\centering
{\includegraphics[width=.9\linewidth]{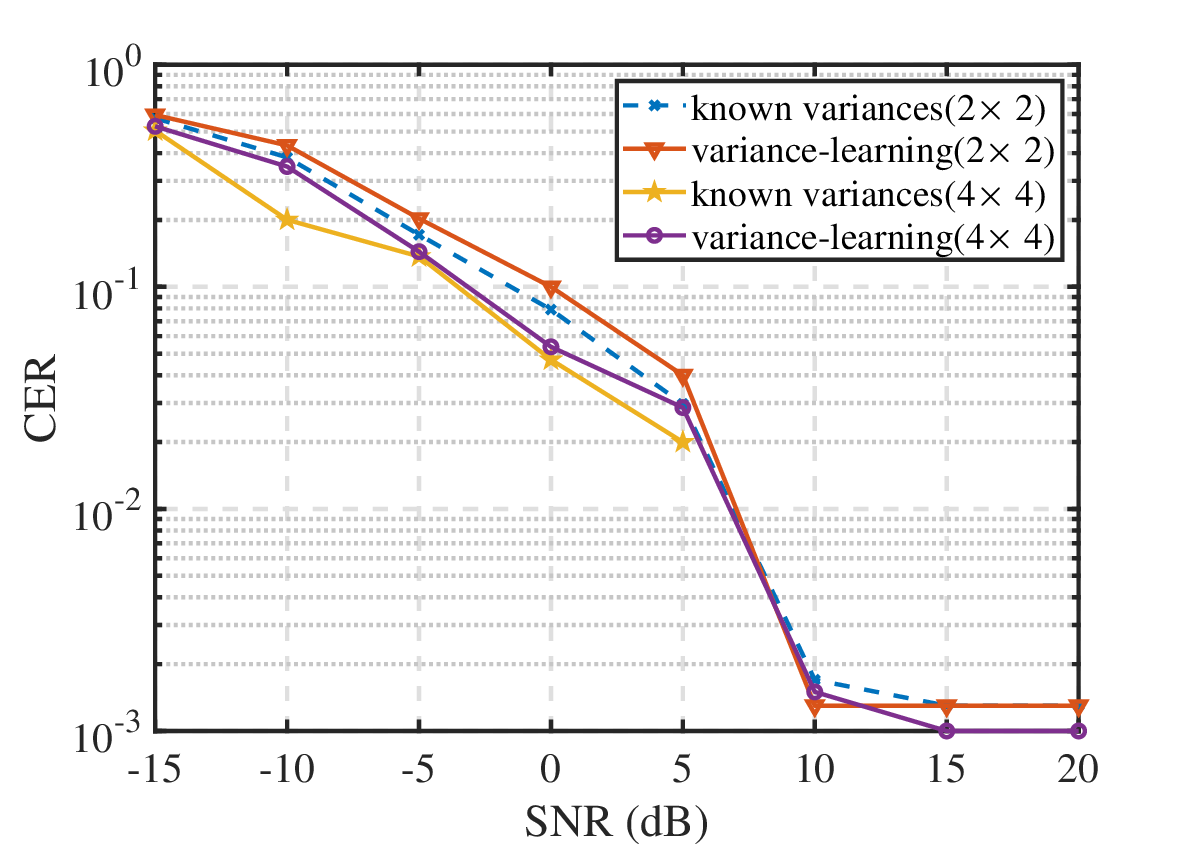}}%
\caption{The CER comparison of the proposed GEM algorithm for two MIMO configurations.}
\label{CER_SNR}
\end{figure}

\noindent\textbf{CER versus SNR:} Fig. \ref{CER_SNR} compares the CER versus SNR for the proposed GEM-based detector under both known-variance and unknown-variance scenarios. The simulations are performed with a block length of $L=40$ and $N_b=3000$ received signal blocks for two MIMO configurations, namely $2\times2$ and $4\times4$. As expected, the CER decreases monotonically as the SNR increases since the observations become more distinguishable in the high-SNR regime. The algorithm with known variance levels provides the best performance because it does not suffer from variance estimation errors. Nevertheless, the proposed joint estimation algorithm achieves a comparable performance, indicating that the variance levels can be accurately learned during the GEM iterations with only a marginal degradation in detection accuracy. Furthermore, the $4\times4$ MIMO configuration consistently outperforms the $2\times2$ case due to the additional spatial diversity and the larger amount of information available for estimating both the channel matrix and the variance states.

\begin{figure}[!t]
\centering
{\includegraphics[width=.9\linewidth]{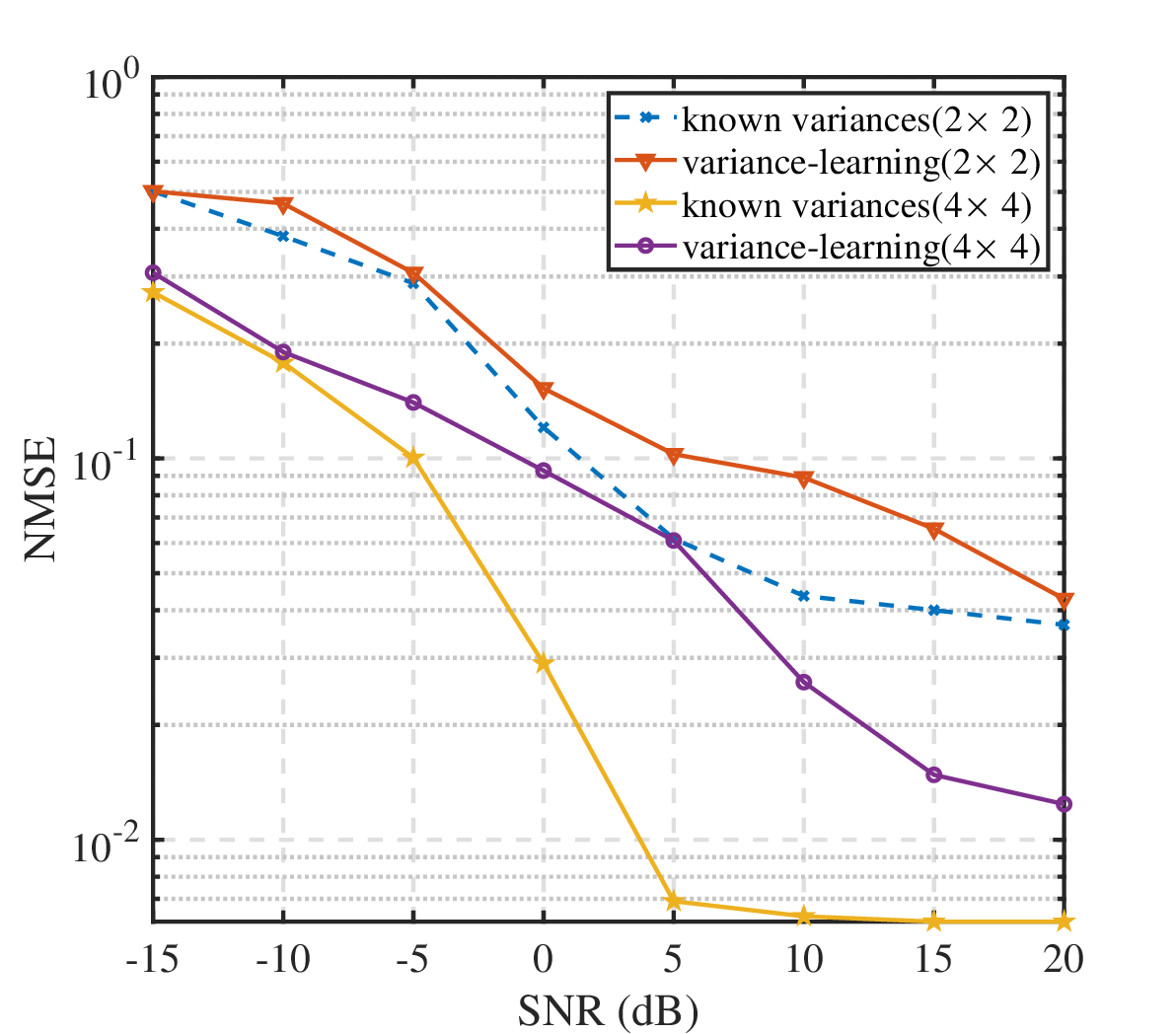}}%
\caption{NMSE comparison of the estimated channel Gram matrix $\mathbf{H}\mathbf{H}^H$ for the proposed GEM algorithm under
two MIMO configurations.}
\label{NMSE_H_SNR}
\end{figure}

\noindent\textbf{Channel NMSE versus SNR:} Fig. \ref{NMSE_H_SNR} illustrates the NMSE of the estimated channel Gram matrix $\mathbf{H}\mathbf{H}^H$ as a function of SNR for both the known-variance and jointly estimated-variance GEM algorithms under the $2\times2$ and $4\times4$ MIMO configurations, with $L=40$ and $N_b=3000$. As the SNR increases, the estimation accuracy of the channel Gram matrix is significantly improved since the effect of additive noise becomes less dominant. As expected, the known-variance algorithm provides the lowest NMSE because the variance parameters are perfectly available during the estimation process. Nevertheless, the proposed joint estimation algorithm closely approaches this performance, demonstrating that accurate estimation of the channel Gram matrix can still be achieved while simultaneously learning the unknown variance levels. Moreover, the $4\times4$ MIMO system generally exhibits improved estimation accuracy due to the increased spatial observations available for the GEM algorithm.

\begin{figure}[!t]
\centering
{\includegraphics[width=.9\linewidth]{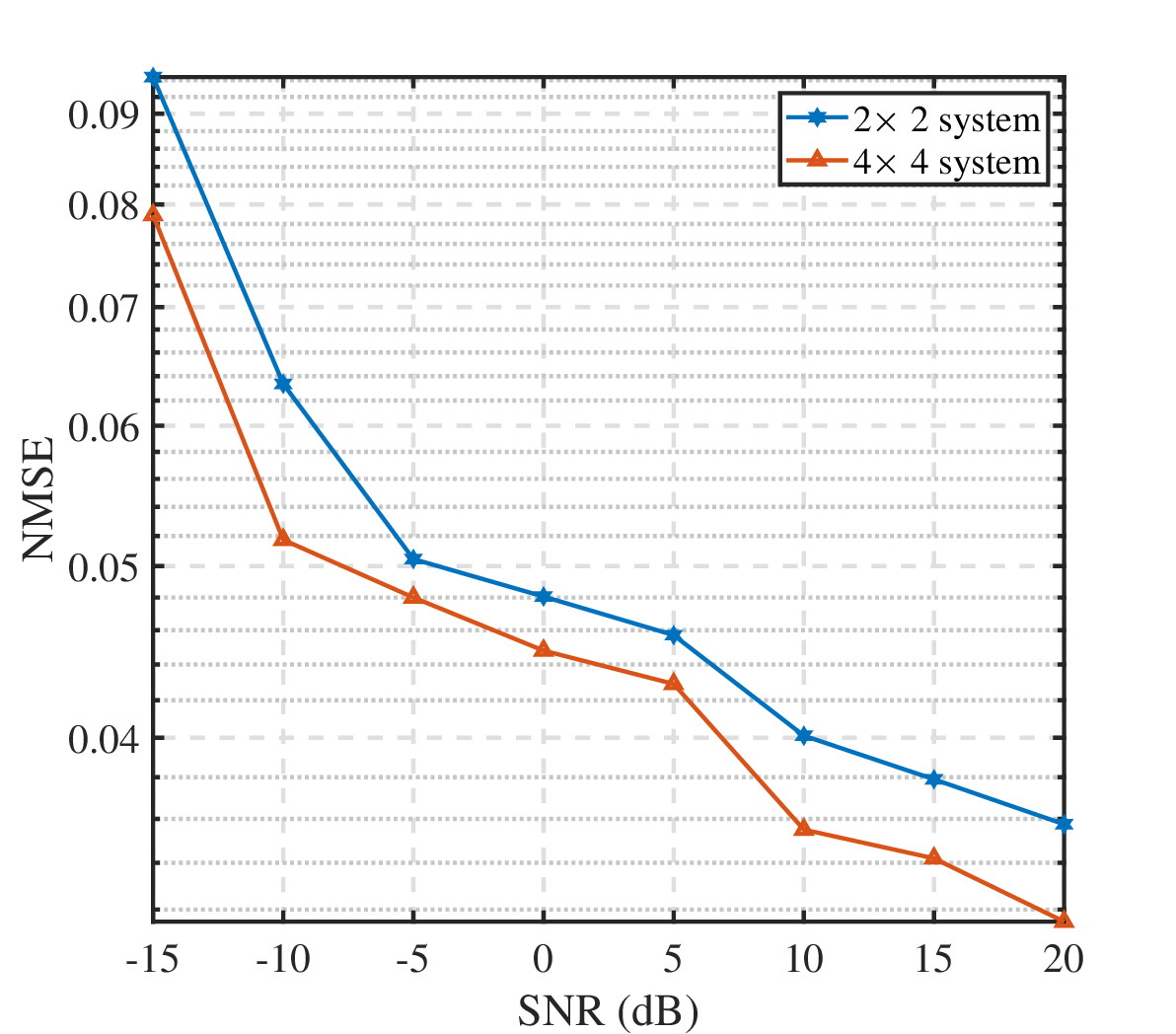}}%
\caption{The NMSE comparison for the variance level estimation of the proposed GEM algorithm for two MIMO configurations.}
\label{NMSE_V_SNR}
\end{figure}

\noindent\textbf{Variance NMSE versus SNR:} Fig. \ref{NMSE_V_SNR} presents the NMSE of the estimated variance levels versus SNR using the proposed GEM algorithm under $2\times2$ and $4\times4$ MIMO systems. As expected, the estimation error decreases with increasing SNR. For instance, at an SNR of 10 dB, the NMSE reaches approximately 0.040 for the $2\times2$ system and 0.036 for the $4\times4$ system. Moreover, the $4\times4$ system consistently achieves lower estimation error than the $2\times2$ system across the considered SNR range. These results demonstrate that the proposed GEM framework can reliably estimate the unknown transmit variance levels directly from the received observations, without prior knowledge of their actual values, and that increasing the number of antennas improves the variance estimation accuracy.

\begin{figure}[!t]
\centering
{\includegraphics[width=.9\linewidth]{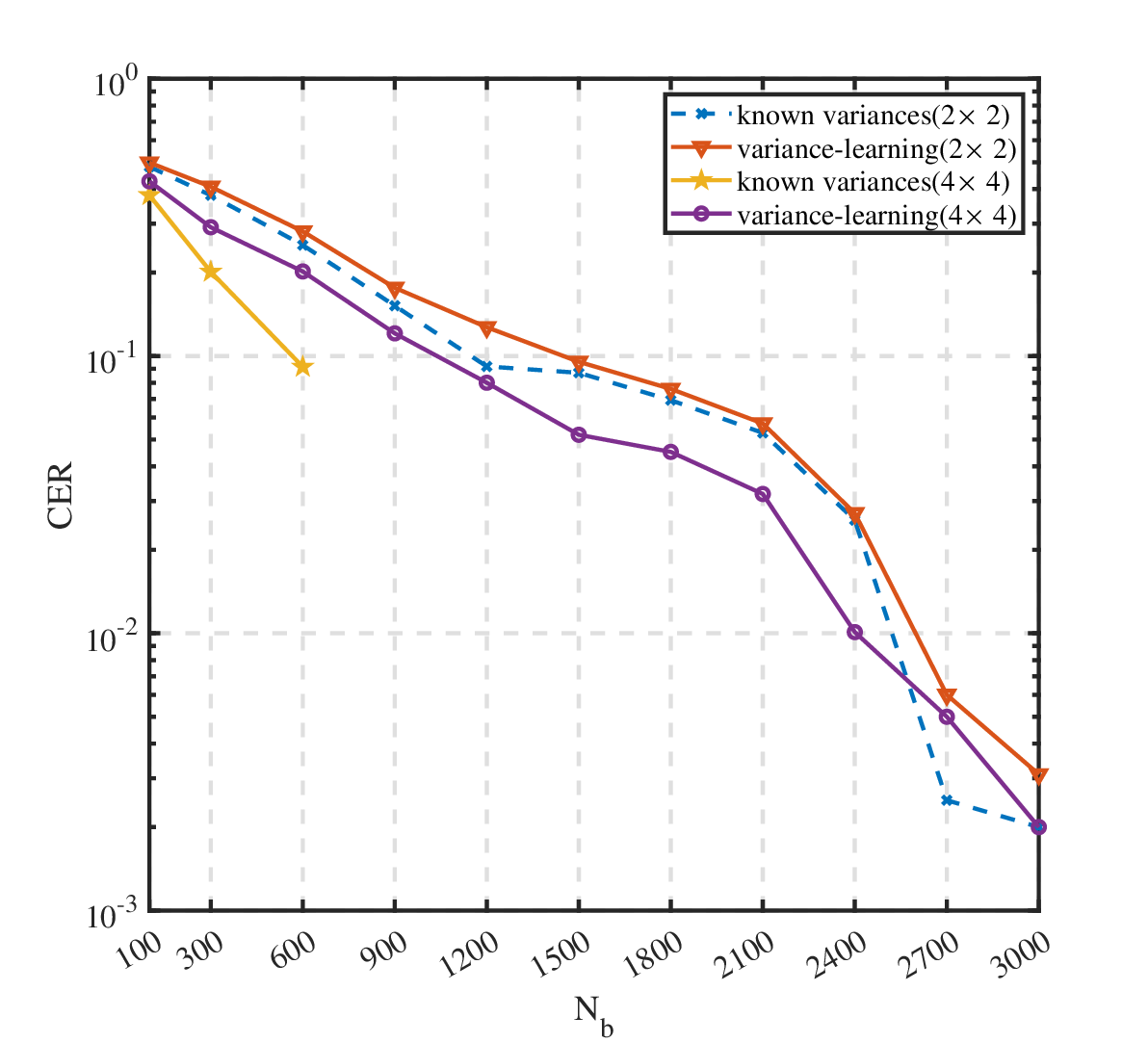}}%
\caption{The CER comparison of the proposed GEM algorithm for different values of $N_b$ under two MIMO configurations at $\mathrm{SNR}=10$ dB.}
\label{CER_K}
\end{figure}

\noindent\textbf{CER versus Number of Observation Blocks:} Fig. \ref{CER_K} illustrates the CER performance of the proposed GEM-based algorithms as a function of the number of received signal blocks $N_b$ at $\mathrm{SNR}=10$ dB. As expected, the CER decreases with increasing $N_b$, since a larger number of observation blocks provides more reliable statistical information for parameter estimation. For instance, in the $2\times2$ system, the CER of the variance-learning algorithm decreases from approximately 0.498 at $N_b=100$ to 0.0031 at $N_b=3000$. Similarly, in the $4\times4$ system, it decreases from 0.427 to 0.002 over the same range. Moreover, the performance gap between the known-variance and variance-learning algorithms becomes smaller as $N_b$ increases. In particular, at $N_b=3000$, both algorithms achieve a CER of approximately 0.002 in the $4\times4$ system. These results demonstrate that the proposed variance-learning procedure becomes increasingly reliable as more observation blocks are available.

\begin{figure}[!t]
\centering
{\includegraphics[width=.9\linewidth]{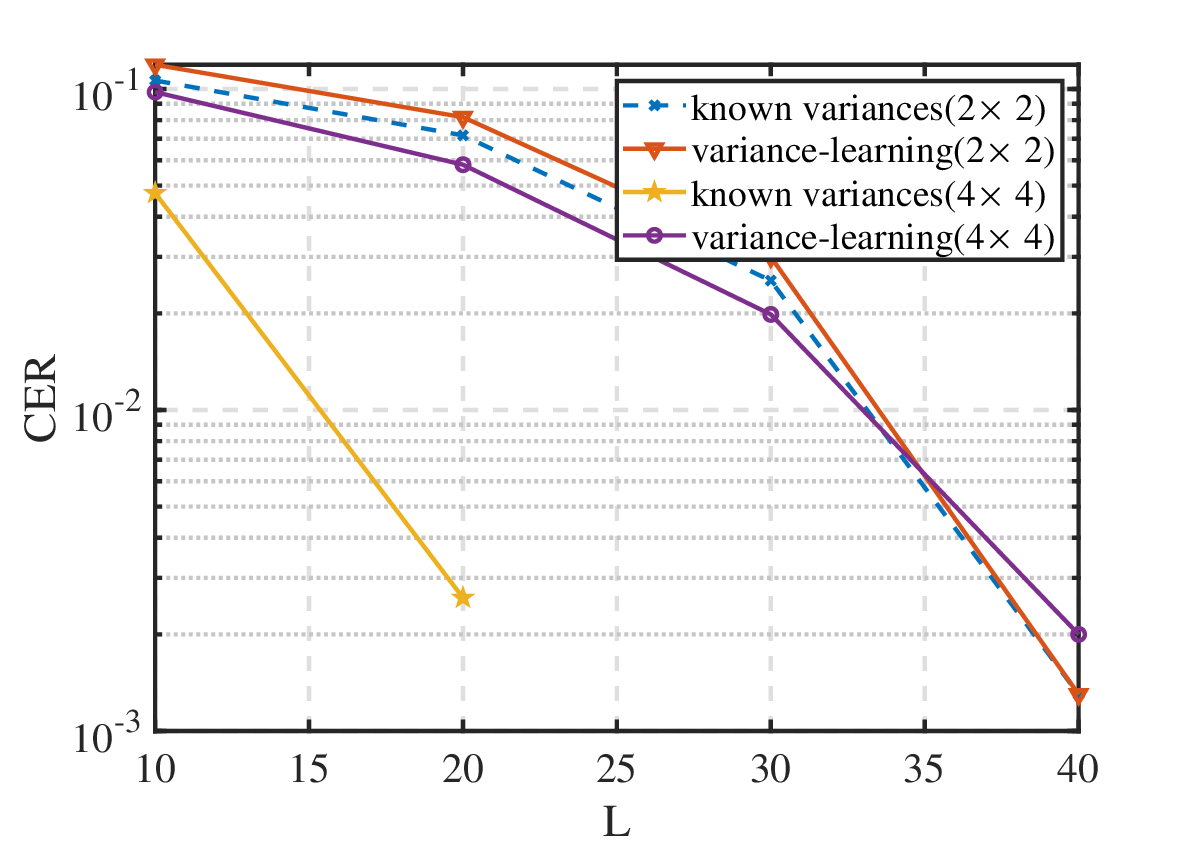}}%
\caption{The CER comparison of the proposed GEM algorithm for different values of $L$ under two MIMO configurations at $\mathrm{SNR}=10$ dB.}
\label{CER_L}
\end{figure}

\noindent\textbf{CER versus Block Length:} Fig. \ref{CER_L} presents the CER performance versus the block length $L$ at $\mathrm{SNR}=10$ dB. It can be observed that increasing the number of noise samples within each signal block significantly improves the classification accuracy. This improvement is attributed to the fact that the sample covariance matrix becomes a more accurate approximation of its theoretical counterpart as $L$ increases, leading to more reliable GEM updates and more accurate estimation of both the channel covariance matrix and the variance levels. Consequently, the CER decreases for both the known-variance and joint estimation algorithms, while the performance difference between the two methods becomes smaller for sufficiently large block lengths.

\begin{figure*}[!t]
\centering
\subfloat[EM objective function]{\includegraphics*[width=0.48\textwidth,height=0.28\textheight]{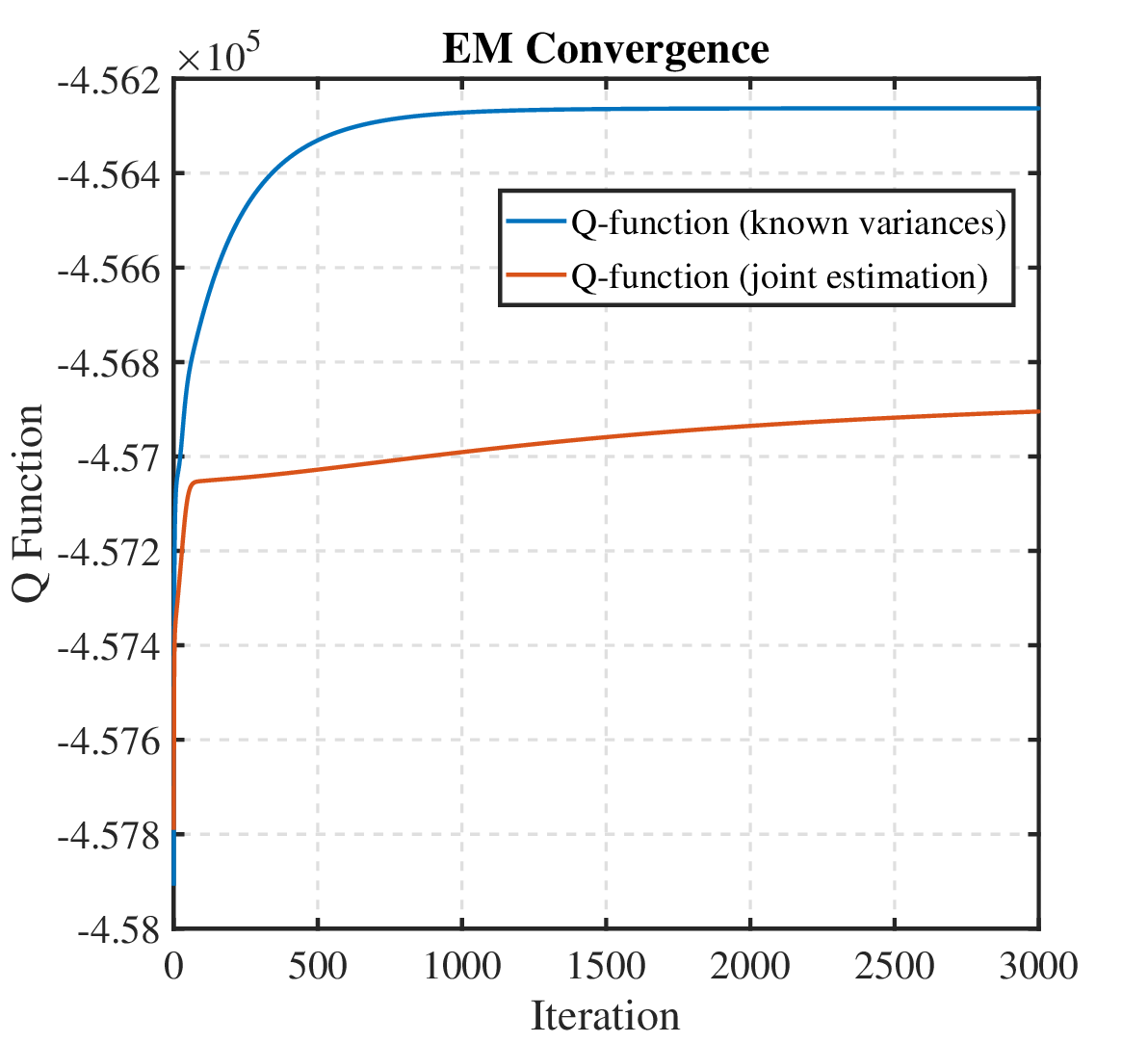}\label{conv} } \hspace*{-0.3em}
\subfloat[variance estimation results]{\includegraphics*[width=0.48\textwidth,height=0.28\textheight]{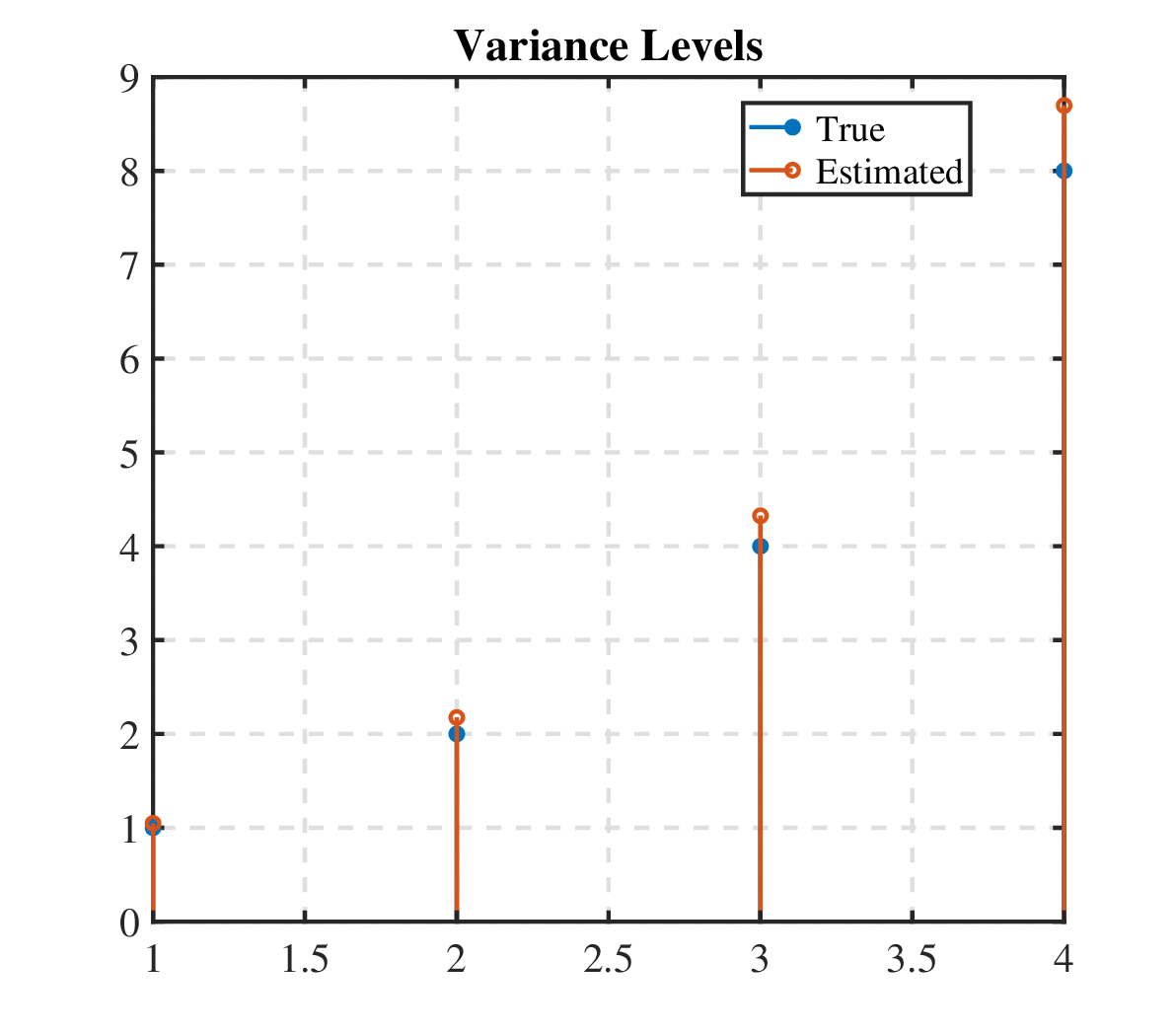}\label{var} }
\caption{Convergence behavior of the GEM algorithm for a $2\times2$ MIMO system: comparison between channel estimation with known variance levels and joint channel and variance-level estimation.}
\label{fig:conv_var}
\end{figure*}

\noindent\textbf{Convergence Behavior:} Fig. \ref{fig:conv_var} presents the convergence behavior of the proposed GEM algorithm and the corresponding variance estimation results for a $2\times2$ MIMO system with $N_b=1500$, $L=40$, and $\mathrm{SNR}=10$ dB. Fig. \ref{conv} illustrates the evolution of the EM objective function with respect to the iteration number for two estimation scenarios. The first curve, denoted by $Q$-function (known variances), corresponds to the case where the true variance levels are assumed to be known and only the channel matrix is estimated. The second curve, denoted by $Q$-function (joint estimation), represents the proposed GEM algorithm, in which both the channel matrix and the variance levels are jointly estimated. As shown in the figure, both curves exhibit a smooth and non-decreasing behavior throughout the iterations, consistent with the convergence analysis presented in Section~\ref{convergence}. The gradient-ascent step sizes are selected empirically through preliminary simulations, guided by the sufficient step-size conditions discussed in Section~\ref{convergence}, to ensure stable convergence while maintaining a reasonable convergence rate. The selected step sizes are then kept fixed throughout the iterations. Fig. \ref{var} compares the true and estimated variance levels using a bar chart under the same simulation conditions. The estimated variance levels closely match their corresponding true values, demonstrating that the proposed joint estimation algorithm can accurately learn the unknown variance states directly from the received observations. These results further confirm the effectiveness of the proposed framework for jointly estimating the channel statistics and the modulation variance levels.

\noindent\textbf{Benchmark Comparison:} Fig.~\ref{compare_MM} compares the BER performance of the proposed blind GEM-based receiver with three benchmark methods, namely the maximum likelihood (ML), variational inference (VI), and linear minimum mean square error (LMMSE) detectors reported in \cite{Yang26}, under the same normalized configuration with $N_b=512$. Both the benchmark algorithms and the proposed receiver are evaluated for the same $4\times8$ MIMO system and employ the same observation length ($L$ received noise samples per TNM symbol). Moreover, the BER is computed using the same symbol-to-bit mapping. As explained at the beginning of this section, the BER curves are plotted versus $E_b/N_0$ using the adopted energy normalization. It should be noted that the proposed receiver does not require pilot symbols, whereas the benchmark methods follow the pilot-assisted configuration described in \cite{Yang26}. Therefore, the proposed approach avoids the pilot-transmission overhead required by the benchmark methods, although this overhead is not explicitly incorporated into the plotted $E_b/N_0$ values.

\begin{figure}[!t]
\centering
{\includegraphics[width=.75\linewidth]{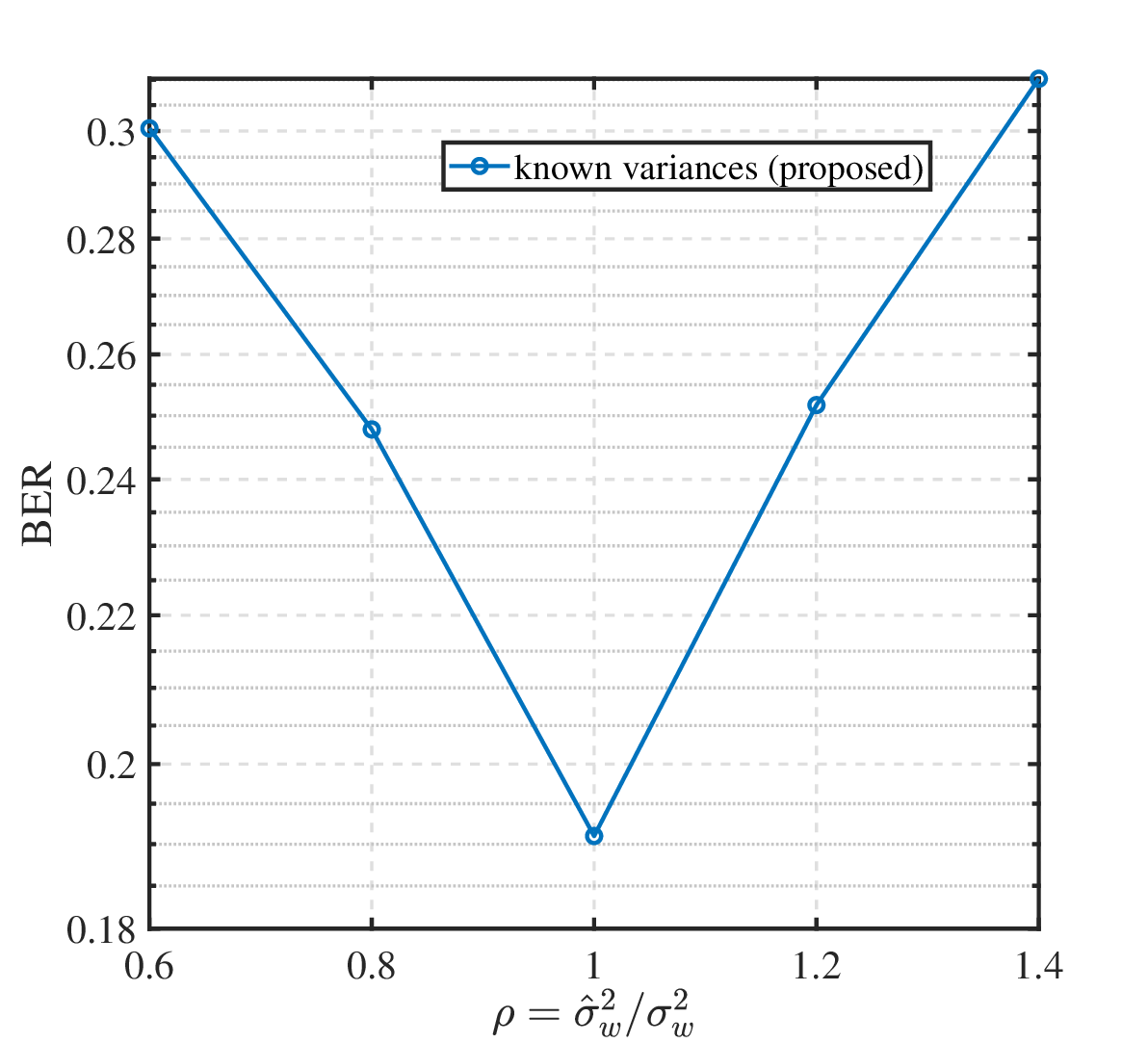}}%
\caption{BER performance of the proposed receiver under receiver-noise variance mismatch ($N_t=N_r=2$, $M=4$, $N_b=1200$, $L=40$, and SNR = 10 dB).}
\label{mismatch}
\end{figure}

\noindent\textbf{Noise Variance Mismatch:} To investigate the sensitivity of the proposed receiver to imperfect receiver-noise variance calibration, additional simulations are conducted by varying the mismatch ratio $\rho=\widehat{\sigma}_w^2/\sigma_w^2$, where $\widehat{\sigma}_w^2$ denotes the noise variance assumed at the receiver. Fig.~\ref{mismatch} shows the BER performance for different values of $\rho$. As expected, the lowest BER is achieved when the noise variance is accurately estimated ($\rho=1$). Both underestimation and overestimation of the receiver-noise variance degrade the detection performance, highlighting the importance of accurate noise variance calibration.

Two variants of the proposed receiver are considered: (i) assuming that the transmit noise variance levels are perfectly known and (ii) jointly estimating both the channel matrix and the transmit noise variance levels. As expected, the ML detector provides the best detection performance since it represents the optimal benchmark. The VI detector achieves near-optimal performance while maintaining significantly lower computational complexity. Compared with these pilot-aided approaches, the proposed blind receiver exhibits a moderate performance degradation due to the absence of pilot symbols and the need to jointly estimate the channel and unknown parameters. Furthermore, the results demonstrate that the proposed receiver with known variance levels consistently outperforms the variance-learning version, indicating the performance loss introduced by blind variance estimation. Nevertheless, the performance gap between the two proposed schemes remains relatively small over the considered SNR range, demonstrating the effectiveness and robustness of the proposed joint estimation framework.

\noindent\textbf{Variance Constellation Generality:} To further assess the generality of the proposed receiver, simulations are conducted using geometric constellations with $M=2,4,$ and $8$, as well as a non-geometric constellation with $M=4$. The corresponding variance levels are listed in Table \ref{tab:constellations}. All simulations are performed for a $2\times2$ MIMO system with $N_b=1200$, $L=40$, and $\mathrm{SNR}=10$ dB. Table \ref{tab:general_M} compares the receiver performance in terms of CER, BER, and variance estimation NMSE. The results demonstrate that the proposed method can accommodate different constellation sizes and structures, including non-geometrically spaced variance levels and constellations with a minimum value different from unity.

\begin{table}[t]
\centering
\caption{Variance constellations considered in the simulations.}
\label{tab:constellations}
\renewcommand{\arraystretch}{1.15}
\begin{tabular}{|c| c| l|}
\hline
$M$ & Type & Variance levels ($\mathcal{V}$) \\
\hline
2 & Geometric & $\{1,4\}$ \\
\hline
4 & Geometric & $\{1,2,4,8\}$ \\
\hline
4 & Non-geometric & $\{0.5,1.3,2.7,5.2\}$ \\
\hline
8 & Geometric & $\{1,1.5,\ldots,17.0859\}$ \\
\hline
\end{tabular}
\end{table}

\begin{table}[t]
\centering
\caption{Performance comparison for different variance constellations at $\mathrm{SNR}=10$ dB.}
\label{tab:general_M}
\renewcommand{\arraystretch}{1.15}
\setlength{\tabcolsep}{4pt}
\begin{tabular}{|c| c| c| c| c| c|}
\hline
$M$ & Type & Receiver & CER & BER & $NMSE_v$ \\
\hline
2 & Geo. & Known   & 0.0010 & 0.0911 & -- \\
2 & Geo. & Learned & 0.0012 & 0.0911 & 0.0134 \\
\hline
4 & Geo. & Known   & 0.0018 & 0.1502 & -- \\
4 & Geo. & Learned & 0.0020 & 0.1671 & 0.0411 \\
4 & Non-geo. & Known   & 0.0020 & 0.1631 & -- \\
4 & Non-geo. & Learned & 0.0017 & 0.1747 & 0.0528 \\
\hline
8 & Geo. & Known   & 0.0026 & 0.1759 & -- \\
8 & Geo. & Learned & 0.0028 & 0.1903 & 0.0691 \\
\hline
\end{tabular}
\end{table}

\noindent\textbf{Non-Gaussian Noise Robustness:} To investigate the robustness of the proposed receiver against non-Gaussian noise, additional simulations are conducted using an impulsive Gaussian-mixture noise model. Specifically, each receiver-noise sample is independently generated from a two-component complex Gaussian mixture: it follows the background Gaussian distribution with probability \(1-p_{\mathrm{imp}}\) and the impulsive Gaussian distribution with probability \(p_{\mathrm{imp}}\). The impulsive-noise variance is set to ten times the background-noise variance. To ensure a fair comparison, the two component variances are adjusted such that the average total receiver-noise variance remains equal to that of the nominal Gaussian-noise scenario. Table \ref{tab:non_gaussian} presents the CER, BER, and variance estimation NMSE. The results show that the performance of both receivers degrades as the impulse probability increases.

\begin{table}[t]
\centering
\caption{Performance of the proposed receivers under Gaussian and non-Gaussian receiver noise.}
\label{tab:non_gaussian}
\renewcommand{\arraystretch}{1.15}
\setlength{\tabcolsep}{5pt}
\begin{tabular}{|c| c| c| c| c|}
\hline
Noise model & Receiver & CER & BER & $NMSE_v$ \\
\hline
\multirow{2}{*}{Gaussian ($\epsilon=0$)}
& Known variance & 0.0018 & 0.1502 & -- \\
& Learned variance & 0.0020 & 0.1671 & 0.0411 \\
\hline
\multirow{2}{*}{Gaussian mixture ($\epsilon=0.01$)}
& Known variance & 0.0019 & 0.1612 & -- \\
& Learned variance & 0.0022 & 0.1892 & 0.0921 \\
\hline
\multirow{2}{*}{Gaussian mixture ($\epsilon=0.05$)}
& Known variance & 0.0031 & 0.2712 & -- \\
& Learned variance & 0.0037 & 0.3224 & 0.1758 \\
\hline
\end{tabular}
\end{table}

\noindent\textbf{Initialization Sensitivity:} To investigate the behavior of the proposed algorithm under different parameter initializations, three different combinations of initial variance levels and mixture probabilities are considered, as listed in Table \ref{tab:initializations}. The corresponding performance results are reported in Table \ref{tab:initialization_results}. The results show that the proposed receiver achieves satisfactory detection performance under all three considered initializations. Although some variations in variance estimation accuracy are observed, with the first initialization yielding the lowest NMSE, the overall detection performance remains relatively consistent across the tested initial conditions. These results provide insight into the sensitivity of the proposed receiver to parameter initialization.

\begin{table}[t]
\centering
\caption{Different initializations of the unknown parameters.}
\label{tab:initializations}
\begin{tabular}{|c|c|c|c|}
\hline
Initialization & $\mathbf{H}^{(0)}$ & $\mathbf{v}^{(0)}$ & $\boldsymbol{\pi}^{(0)}$ \\
\hline
1 & $0.5\mathbf{H}_0$ & $[0.2,\,0.8,\,2,\,5]$ & $[0.25,\,0.25,\,0.25,\,0.25]$ \\
\hline
2 & $\mathbf{H}_0$ & $[0.3,\,1.5,\,3,\,10]$ & $[0.1,\,0.2,\,0.3,\,0.4]$ \\
\hline
3 & $2\mathbf{H}_0$ & $[1,\,3,\,6,\,12]$ & $[0.4,\,0.3,\,0.2,\,0.1]$ \\
\hline
\end{tabular}
\end{table}

\begin{table}[t]
\centering
\caption{Performance of the proposed receiver under different initializations.}
\label{tab:initialization_results}
\renewcommand{\arraystretch}{1.2}
\setlength{\tabcolsep}{7pt}
\begin{tabular}{|c|c|c|c|c|}
\hline
Initialization & Receiver & CER & BER & NMSE$_v$ \\
\hline
\multirow{2}{*}{1}
& Known variance & 0.0018 & 0.1502 & -- \\
& Learned variance & 0.0020 & 0.1671 & 0.0411 \\
\hline
\multirow{2}{*}{2}
& Known variance & 0.0021 & 0.1673 & -- \\
& Learned variance & 0.0024 & 0.1947 & 0.0988 \\
\hline
\multirow{2}{*}{3}
& Known variance & 0.0022 & 0.1714 & -- \\
& Learned variance & 0.0027 & 0.1994 & 0.1048 \\
\hline
\end{tabular}
\end{table}

\section{Conclusion}\label{conclusion}

This paper proposed a pilot-free blind receiver based on the Generalized Expectation-Maximization (GEM) algorithm for M-ary thermal noise modulation (TNM) systems. By exploiting the statistical characteristics of TNM signals, the proposed framework jointly estimates the channel matrix and unknown transmit noise variance levels, followed by symbol detection, without requiring pilot transmission. Consequently, the proposed receiver completely eliminates pilot overhead while providing reliable channel estimation and symbol detection.

Simulation results demonstrate that the proposed blind receiver achieves competitive BER performance compared with existing pilot-aided benchmark algorithms. Although the joint estimation of the transmit noise variance levels introduces a moderate performance degradation compared with the ideal case of perfectly known variances, the resulting performance loss remains relatively small over the entire SNR range considered. Furthermore, the proposed variance-learning receiver significantly outperforms the conventional LMMSE detector and achieves performance close to the variational inference (VI) detector while operating in a completely blind manner. These results verify the effectiveness and robustness of the proposed joint estimation framework and demonstrate its suitability for practical TNM systems where neither pilot symbols nor prior knowledge of the variance constellation is available.

Despite these promising results, the proposed approach has some limitations. Its estimation performance may depend on the initialization of the channel matrix and variance levels, while the gradient-based updates require appropriate step-size selection to ensure stable convergence. Moreover, owing to the non-convex nature of the underlying optimization problem, the algorithm may converge to a local stationary point rather than the global optimum. The current formulation also assumes that the receiver-noise variance is known, and inaccuracies in this parameter may affect the estimation accuracy. These limitations suggest several directions for future research, including developing more robust initialization strategies, designing adaptive step-size selection methods, investigating approaches to improve convergence toward better solutions, and extending the proposed framework to jointly estimate the receiver-noise variance along with the channel matrix and transmit variance levels.

Future work may consider extending the proposed framework to more general receiver-noise models, including spatially correlated noise and unequal noise variances across receive antennas. Such extensions would require a generalized receiver-noise covariance model and corresponding modifications to the estimation algorithm.

\end{document}